\documentclass[seceq]{ptptex}
\usepackage{graphicx}
\usepackage{amssymb,amsmath}
\nofigureboxrule                     
\notypesetlogo                       

\title{
Thermodynamics of Quantum Ultra-cold Neutron Gas under Gravity of The Earth%
}

\author{
Hiromi \textsc{Kaneko}\footnote{kanekoh@rcnp.osaka-u.ac.jp},
Akihiro \textsc{Tohsaki}\footnote{tohsaki@rcnp.osaka-u.ac.jp} and 
Atsushi \textsc{Hosaka}\footnote{hosaka@rcnp.osaka-u.ac.jp}
}

\inst{
Research Center for Nuclear Physics, Osaka University, Mihogaoka 10-1, Osaka 567-0047, Japan
}

\abst{
The stored ultra-cold neutrons have been developed.
A high density ultra-cold neutron gas has been recently produced by using the nuclear spallation method.
We investigate the thermodynamic properties of the quantum ultra-cold neutron gas in the Earth's gravitational field.
We find that the quantum effects increase temperature dependence of the chemical potential and the internal energy in the low temperature region.
The density distribution of quantum ultra-cold neutron gas is modified by the Earth's gravitational field.
}
\begin{document}
\maketitle
\section{Introduction}
\label{sec.1}
Neutrons are difficult to trap in any container because of their neutral nature.
Very slow neutrons, however, are found to be reflected by a metallic wall, and therefore, can be stored in metallic containers; see Ref.~\citen{Golub:1979zz} and references therein.
They are the so-called Ultra-cold Neutrons (UCNs).
The technology of producing UCN has been developed in order to search for the time reversal violating electric dipole moment of the neutron and for the most accurate measurement of the neutron $\beta$ decay lifetime.
The UCN scattering is applied to the study of condensed matter properties~\citen{Golub:1996zz}. 
Recent experiments have succeeded in creating UCN gas by using nuclear spallation method,
where the neutron density is not limited by Liouville's theorem~\citen{PhysRevLett.108.134801,Masuda:2002dy,Saunders:2003jg,Atchison:2005fr}.
The density of UCNs coming from a nuclear reactor never increases under this restriction.
Therefore the nuclear spallation method is expected to create high density UCN gas and also to study quantum neutron systems.
\par
Turning to nuclear physics, neutron systems have been studied to discover interesting phenomena originating in the strong interaction like the neutron star~\citen{Glendenning200003}.
This is a huge quantum system whose typical neutron density is of the order of $1$~fm$^{-3}$ (=$10^{39}$~cm$^{-3}$). 
In the current experimental state~\citen{PhysRevLett.108.134801} the UCN gas is very dilute, with density $\sim 10^2$~cm$^{-3}$,
At such low densities, thermodynamic properties are described by the Maxwell-Boltzmann distribution:
the mean separation length between neutrons is $\bar{R} = (\text{density})^{-1/3}\sim 2\times 10^{-1}$~cm, 
the thermal de Broglie wave-length is $\bar{\lambda} = h/(3mk_BT)^{1/2} \sim 8\times 10^{-6}$~cm at $1$~mK,
where $m$ is neutron mass, $k_B$ is the Boltzmann constant and $T$ is the temperature.
Due to the very long wave-length, the UCN can be reflected and stored in metallic containers.
\par
In this paper, we investigate the thermodynamic properties of the quantum neutron gas in the Earth's gravitational field.
With developing UCN experiments to create high density or low temperature gas,
a treatment of quantum statistical mechanics is needed.
The neutron density is expected to increase under the influence of the gravity, which is
even more enhanced as the temperature becomes lower.
There the chemical potential and the internal energy increase as compared to the case of
the free Fermi gas.
The density is the highest at the bottom of the container, and it becomes smaller with increasing height and with increasing temperature. 
In Sec.~\ref{sec.2} we show the wave function and the eigenenergies of a neutron trapped by the Earth's gravitational field.
Secs.~\ref{sec.3} and \ref{sec.4} are devoted to the internal energy and the density distribution by using the chemical potential.
Finally, we conclude the present study in Sec.~\ref{sec.5}.
\section{Neutron in the Earth's Gravitational Field}
\label{sec.2}
Let us first consider the one-neutron wave function trapped by the Earth's gravitational field.
The Shr\"{o}dinger equation reads
\begin{equation}
-\frac{\hbar^2}{2m}\frac{d^2\psi(z)}{dz^2}+mgz\psi(z)  = E_z\psi(z)\, ,
\label{eq.shr_z}
\end{equation}
where $g$ is the standard gravitational acceleration, the $z$-axis is along with the Earth's gravitational field and $z=0$ is the bottom of the container.
By introducing  $\alpha = 2m^2g/\hbar^2$, $\epsilon = (\alpha/mg)E_z$ and $\xi = (\alpha z - \epsilon)/\alpha^{2/3}$,
Eq.~(\ref{eq.shr_z}) is rewritten as a dimensionless Schr\"{o}dinger equation
\begin{equation}
 \frac{d^2\psi(\xi)}{d\xi^2}-\xi\psi(\xi) = 0 \,.
\end{equation}
The general solutions are given by the well-known Airy functions (see Ref.~\citen{Abramowitz_Stegun196506}) 
\begin{equation}
\psi(\xi) = N_aAi(\xi) +N_bBi(\xi)\, ,
\end{equation}
where $N_a$ and $N_b$ determine the weight of the two independent functions.
Assuming the boundary condition that $\lim_{z\to\infty}\psi(z)=0$, the factor of $N_b$ should be zero, because $Bi(\xi)$ diverges at infinity.
Therefore the wave function is proportional to $Ai(\xi)$, 
\begin{equation}
\psi_{n_z}(z)  = N_aAi\left(\frac{\alpha z - \epsilon_{n_z}}{\alpha^{2/3}}\right)\, .
\label{eq.wavefunction}
\end{equation}
By imposing another boundary condition $\psi_{n_z}(0)=0$, the wave function must obey
\begin{equation}
 Ai\left(-\frac{\epsilon_{n_z}}{\alpha^{2/3}}\right)=0\, ,
\end{equation}
which determines the eigenenergies because the neutron is reflected by the bottom of the container.
By using the asymptotic behavior of zeros of the Airy functions~\citen{Abramowitz_Stegun196506},
the eigenenergies
\footnote{This is a good approximation at 1~\% or better}
are given by
\begin{equation}
 E_{n_z} \approx \left(\frac{mg^2\hbar^2}{2}\right)^{1/3}\left(\frac{3\pi}{8}\right)^{2/3}(4n_z-1)^{2/3}\, ,\, n_z = 1,2\ldots\, . 
\label{eq.energy_z}
\end{equation}
Substituting $g=0$ into Eq.~(\ref{eq.energy_z}), obviously, the energies are vanished.
This reflects the fact that the $z$-direction is not bounded.
The normalization $N_a$ in Eq.~(\ref{eq.wavefunction}) is determined from the orthonormalization condition of the wave functions, 
\begin{equation}
 \int^{\infty}_0 \psi_n(z)\psi_m^*(z) dz = \delta_{n,m}\, .
\end{equation}
By using the integration formula from Ref.~\citen{ValleeSoares201008}, the normalized wave functions are given by    
\begin{equation}
\psi_{n_z}(z) = \frac{\alpha^{1/6}}{|Ai'(-\epsilon_{n_z}/\alpha^{2/3})|}Ai\left(\frac{\alpha z-\epsilon_{n_z}}{\alpha^{2/3}}\right)\, .
\end{equation} 
Furthermore the very slow neutron is reflected by a metallic wall in $x$- and $y$-axes.
For simplicity we consider a square-well potential of infinite height.
Thus the energies are $E_{n}=\pi^2\hbar^2n^2/2mL^2$, where $L$ is the size of square-well and $n$ is a positive integer.
Therefore, the total energies are written by the sum with respect to the three-axes: 
\begin{align}
E_{n_x,n_y,n_z} = \frac{\pi^2\hbar^2}{2mL^2}(n_x^2+n_y^2)+\left(\frac{mg^2\hbar^2}{2}\right)^{1/3}\left(\frac{3\pi}{8}\right)^{2/3}(4n_z-1)^{2/3}\, ,
\label{eq.totalenergy}
\end{align}
where $n_x$, $n_y$ and $n_z$ are arbitrary positive integers.
\section{Density of States, Chemical Potential and Internal Energy}
\label{sec.3}
\begin{figure}[htbp]
\begin{center}
\includegraphics[width = 0.47\textwidth]{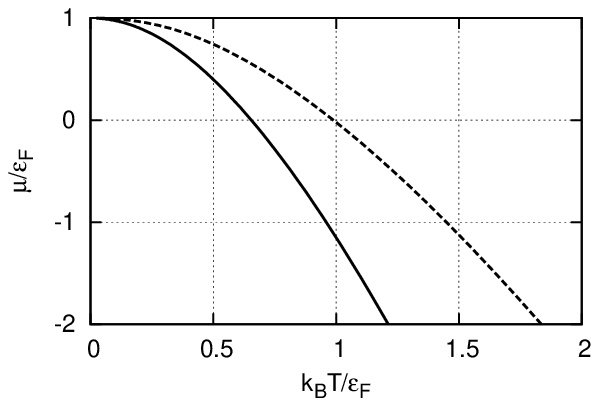}
\includegraphics[width = 0.47\textwidth]{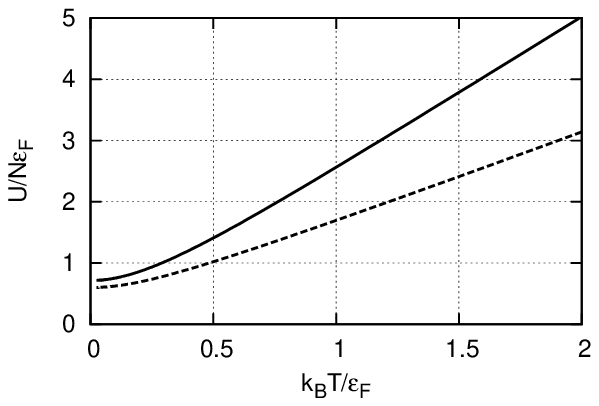}
\caption{The chemical potential and the  internal energy as functions of the temperatures: the solid lines present the results of Fermi gas under the Earth's gravitational field, the dashed lines are those of the free Fermi gas.}
\label{fig.chem_int}			
\end{center}
\end{figure}

To derive the thermodynamic properties,
let us calculate the density of states by using Eq.~(\ref{eq.totalenergy}):
\begin{equation}
\rho(E_{n_x},E_{n_y},E_{n_z})
=\frac{dn_x}{dE_{n_x}}\frac{dn_y}{dE_{n_y}}\frac{dn_z}{dE_{n_z}} 
= \frac{2m}{\hbar^2}\left(\frac{L}{2\pi}\right)^2 \frac{1}{\pi}\sqrt{\frac{2}{mg^2\hbar^2}}\sqrt{\frac{E_{n_z}}{E_{n_x}E_{n_y}}}\, .
\label{eq.sd}
\end{equation}
The chemical potential $\mu$ can be determined from the following condition:
\begin{equation}
 N = 2\int_0^{\infty} dE_{n_x}dE_{n_y} dE_{n_z} \frac{\rho(E_{n_x},E_{n_y},E_{n_z})}{e^{\beta(E_{n_x}+E_{n_y}+E_{n_z}-\mu)}+1}\, ,
\label{eq.condition}
\end{equation}
where $N$ is the total number of particles and the factor of 2 comes from spin degeneracy.
The coefficient $\beta$ is defined as $(k_BT)^{-1}$, where $k_B$ is the Boltzmann constant and $T$ is the temperature.
In the limit $T \to 0$ the Fermi energy is
\begin{equation}
\epsilon_F = \frac{\hbar^2}{2m} \left( \frac{15\pi^2m^2g}{\hbar^2}\frac{N}{L^2}\right)^{2/5}\, . 
\label{eq.fermienergy}
\end{equation}
Notice that it is different from that of the free Fermi gas; the volume depends on the Earth's gravitational field; the power in Eq.~(\ref{eq.fermienergy}) is $2/5$, which is related with the density. 
To simplify the density of states, putting $E_{n_r} = E_{n_x}+E_{n_y}$ ($n_x=n_r\cos\theta$ and $n_y=n_r\sin\theta$) in Eq.~(\ref{eq.condition}), it is rewritten down
\begin{equation}
 N  = \frac{15N}{8\epsilon_F^{5/2}}\int_0^{\infty}\int_0^{\infty}\frac{\sqrt{E_{n_z}}}{e^{\beta(E_{n_r}+E_{n_z}-\mu)}}dE_{n_r}dE_{n_z}\, ,
\label{eq.redefinedconditon}
\end{equation}
where we use $\epsilon_F$ and
\begin{equation}
 \int_0^{\infty}\int_0^{\infty}\frac{dE_{n_x}dE_{n_y}}{\sqrt{E_{n_x}E_{n_y}}} = \frac{2\hbar^2\pi^2}{mL^2}\int_0^{\infty}\int_0^{\infty}dn_xdn_y  =  \int_0^{\pi/2}d\theta\int_0^{\infty} \frac{2\hbar^2\pi^2n_r}{mL^2}d{n_r}=\pi\int_0^{\infty}dE_{n_r} \, .
\end{equation}
The density of states can be redefined as
\begin{equation}
 \rho(E_{n_r}, E_{n_z}) = \frac{15N}{8\epsilon_F^{5/2}}\sqrt{E_{n_z}}\, .
\label{eq.redefined_density_of_states}
\end{equation}
Setting $\eta = \beta\mu$, $\zeta = \beta E_{n_z}$ and $\upsilon = \beta E_{n_r}$ for Eq.~(\ref{eq.redefinedconditon}),
$\beta$ is regarded as a function of $\eta$,
\begin{equation}
\beta\epsilon_F  = \left(\frac{15}{4}\int_0^{\infty}\int_0^{\infty}\frac{\zeta^{1/2}}{e^{\zeta+\upsilon-\eta}+1}d\zeta d\upsilon\right)^{2/5} = \left(\frac{5}{2}\int_0^{\infty}\frac{\zeta^{3/2}}{e^{\zeta-\eta}+1}d\zeta\right)^{2/5}\, ,
\label{eq.tem} 
\end{equation}
where the integration over $\upsilon$ is performed by integrated by parts.
Therefore the chemical potential is given by 
\begin{equation} 
\frac{\mu}{\epsilon_F} =  \frac{\eta}{\beta\epsilon_F} = \eta\left(\frac{5}{2}\int_0^{\infty}\frac{\zeta^{3/2}}{e^{\zeta-\eta}+1}d\zeta\right)^{-2/5}\, , 
\label{eq.chem}
\end{equation}
which is normalized by $\epsilon_F$.
In the low temperature approximation, the integration of Eq.~(\ref{eq.chem}) is analyzed in Refs.~\citen{JApplPhys_73_7030,Jmathphys421860},
\begin{equation}
 \int_0^{\infty}\frac{\zeta^{3/2}}{e^{\zeta-\eta}+1}d\zeta =  \frac{2}{5}\eta^{5/2}+\frac{\pi^2}{2}\eta^{1/2}+\ldots\,\,  ,\text{ for}\,\, e^{-\eta} \ll 1\, .
\label{eq.formula.1}
\end{equation}
Thus the chemical potential can be calculated as   
\begin{equation}
\frac{\mu}{\epsilon_F} = 1-\frac{\pi^2}{2}\left(\frac{k_BT}{\epsilon_F}\right)^{2}+\ldots\,\,  ,\text{ for}\,\, e^{-\mu/k_BT} \ll 1\, .  
\end{equation}
The internal energy is given by 
\begin{equation}
\begin{gathered}
U = 2\int_0^{\infty}\int_0^{\infty}(E_{n_r}+E_{n_z})\frac{\rho(E_{n_r},E_{n_z})}{e^{\beta(E+E_{n_z})}+1}dE_{n_r}dE_{n_z} =\\
  = \frac{15}{4}{N\epsilon_F}(\epsilon_F\beta)^{-7/2}\left(\frac{2}{5}\int_0^{\infty}\frac{\zeta^{5/2}}{e^{\zeta-\eta}+1}d\zeta
    +\int_0^{\infty}\int_0^{\infty}\frac{\zeta^{1/2}\upsilon}{e^{\zeta+\upsilon-\eta}+1}d\zeta d\upsilon\right)\, .
\label{eq.int}
\end{gathered}
\end{equation}
In Fig.~\ref{fig.chem_int}, we plot the temperature dependence of the chemical potential and the internal energy using Eqs.~(\ref{eq.tem}, \ref{eq.chem}, \ref{eq.int}),
which are compared with those of the free Fermi gas.
The temperature dependence of the chemical potential is $6$ times larger than that of the free Fermi gas with the same Fermi energy in the low temperature region.
The internal energy also has strong temperature dependence at the low temperature region.
As in the case of the classical gas like atmosphere, this means that the specific heat of quantum UCN gas increases by the Earth's gravitational field. 
\section{Density Distribution}
\label{sec.4}
\begin{figure}[htbp]
 \begin{center}
  \includegraphics[width = 0.6\textwidth]{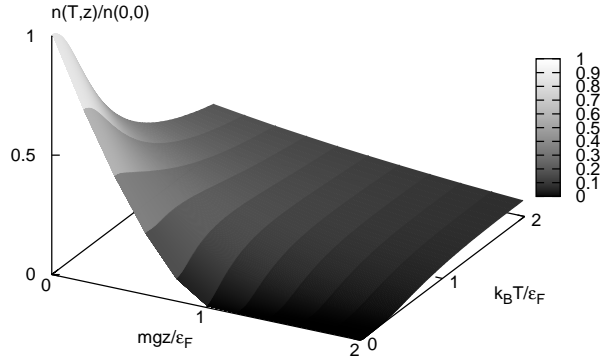}
  \caption{The density distribution as functions of the temperature $k_BT/\epsilon_F$ and $mgz/\epsilon_F$}
  \label{fig.density_3d}
 \end{center}
\end{figure}

\begin{figure}[htbp]
 \begin{center}
  \includegraphics[width = 0.6\textwidth]{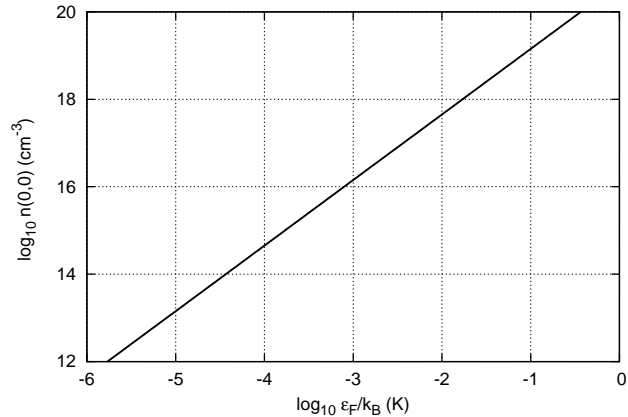}
  \caption{The density at the bottom of the container as a function of the Fermi energy $\epsilon_F/k_B$}
  \label{fig.density_kelvin}
 \end{center}
\end{figure}

It is difficult to observe, for example, the specific heat of UCN gas because the external parameters of pressure or temperature are not easily changed in experiments.
However the density distribution can provide information on the thermodynamic properties, and it can be observed by neutron counting experiments. 
To clearly show the $z$- and $T$- dependencies of the density distribution, following an example shown in Ref~.\citen{RevModPhys.80.1215},
the condition of chemical potential in Eq.~(\ref{eq.condition}) can be rewritten as    
\begin{equation}
N = \frac{1}{(2\pi\hbar)^3}\int d^3xd^3p\frac{1}{e^{(\vec{p}^2/2m+mgz-\mu)/k_BT}+1} = \int_0^{\infty}\frac{\rho(E_{n_r},E_{n_z})}{e^{\beta(E_{n_r}+E_{n_z}-\mu)}+1}dE_{n_r}dE_{n_z}\, .
\label{eq.semicondition}
\end{equation}
Here we define the single particle density as  
\begin{equation}
n(T, z)= \frac{1}{(2\pi\hbar)^3}\int d^3p \frac{1}{e^{(\vec{p}^2/2m+mgz-\mu)/k_BT}+1}\, ,
\label{eq.density}
\end{equation}
then Eq.~(\ref{eq.semicondition}) becomes  
\begin{equation}
N = \int_0^{L}dx\int_0^{L}dy\int_0^{\infty}dz\, n(T,z)\, . 
\end{equation}
Eq.~(\ref{eq.density}) can be calculated in the limit $T \to 0$ as 
\begin{equation}
n(0,z) =  \frac{\left(2m(\epsilon_F-mgz)\right)^{3/2}}{6\pi^2\hbar^3}\,\, ,\text{ for} \,\, \epsilon_F > mgz\, ,
\end{equation}
while it becomes zero for $\epsilon_F < mgz$.
Similarly, putting $\chi = \beta p^2/2m$, the $T$-dependence of the density distribution at $z=0$ is given by
\begin{equation}
\frac{n(T, 0)}{n(0,0)} = \frac{3}{2}\left(\epsilon_F\beta\right)^{-3/2}\int_0^{\infty}  \frac{\chi^{1/2}}{e^{\chi-\eta}+1} d\chi \, , 
\label{eq.normalizeddensity}
\end{equation}
which is normalized by $n(0,0)$.
In the low temperature approximation, the integration is analyzed in Refs.~\citen{JApplPhys_73_7030,Jmathphys421860},
\begin{equation}
 \int_0^{\infty}  \frac{\chi^{1/2}}{e^{\chi-\eta}+1} d\chi = \frac{2}{3}\eta^{3/2} + \frac{\pi^2}{12}\eta^{1/2} \ldots \,\, ,\text{for} \,\, e^{-\eta} \ll 1\, .
\label{eq.formula.2}
\end{equation}
Thus the density distribution at $z=0$ is calculated as
\begin{equation}
 \frac{n(T, 0)}{n(0,0)} = 1-\frac{5\pi^2}{8}\left( \frac{k_B T}{\epsilon_F}\right)^{2} + \ldots \,\, ,\text{ for} \,\, e^{-\mu/k_BT} \ll 1\, ,
\end{equation}
by using the definition of $\beta$ in Eqs.~(\ref{eq.tem}, \ref{eq.formula.1}).
In Fig.~\ref{fig.density_3d}, we plot the density distribution $n(T,z)$ using Eqs.~(\ref{eq.tem}, \ref{eq.density}).
The density distribution with respect to $z$ coordinate rapidly drops with $mgz/\epsilon_F$ in the low temperature region, and it also becomes flat in the high temperature region. 
Finally, we discuss the properties of quantum UCN gas.
In Fig.~\ref{fig.density_kelvin}, we plot the density at the bottom of the container as a function of $\epsilon_F/k_B$ in terms of the double logarithmic form.
For example, when the Fermi energy of quantum UCN gas is $\epsilon_F/k_B\sim 1$~mK, the density becomes $\sim 10^{16}$~cm$^{-3}$ at the bottom and the height also becomes $\epsilon_F/mg\sim 8\times10^1$~cm. In fact, the mean separation length $\bar{R}\sim 5\times 10^{-6}$~cm is close to the thermal de Broglie wave-length $\bar{\lambda} \sim 8\times 10^{-6}$~cm.

\section{Conclusions}
\label{sec.5}
The previous sections provide general thermodynamics properties and the density distribution of the neutron gas under the Earth's gravitational field.
The chemical potential and the internal energy have strong temperature dependence at the low temperatures because the density of sates depends on the square root of the eigen-energies in the Earth' gravitational field.
Thus it can be expected that the variation of specific heat with temperature is larger than that of free Fermi gas.
Instead of the thermodynamic indexes, the density distribution shows how the quantum UCN gas is affected by the Earth's gravitational field.
The density is the highest at the bottom of the container and it becomes small with increasing $mgz/\epsilon_F$ and temperature $T$. 
Assuming that the average kinetic energy of UCNs becomes much smaller than the Fermi energy, the quantum UCN gas would be created.
\section*{Acknowledgments}
We would like to thank M.~Valverde, K.~Nawa, K.~Hatanaka and I.~Tanihata for advice and helpful suggestions.
One of the authours (H.~K.) would like to thank H.~Toki for encouragement.

\end{document}